\documentstyle[12pt]{article}
\begin{document} 

\begin{center} 

{\bf \Large 
Similarity of multifragmentation of Au and Kr residual nucleus}
\end{center} 


\begin{center} 
A.Abd-Elhafiez$^1$, M.M.Chernyavski$^3$, K.G.Gulamov$^2$, 
 V.Sh.Navotny$^2$, G.I.Orlova$^3$, V.V.Uzhinskii$^4$. 
\end{center}

\vskip 1 cm

Experimental data on multi-fragmentation of residual krypton nuclei
created in the interactions of the krypton nuclei with photoemulsion
nuclei at the energy of 0.9 GeV per nucleon are presented in a
comparison with the analogous data on fragmentation of gold residual
nuclei at the energy of 10.7 GeV/nucleon. It is shown for the first
time that there are two regimes of nuclear multifragmentation: the
former is when less than one-half of nucleons of projectile nucleus are
knocked out, the latter is when more than one-half of nucleons are
knocked out. Residual nuclei with closed masses created at different
reactions are fragmenting practically simultaneously when more than
one-half of nucleons of original nuclei are knocked out. The evidence
of existence of a radial flow of the spectator fragment at the decay of
residual krypton nuclei is found.

\vskip 5 cm

\noindent
$^1$ Atomic Energy Authority, Nuclear Research Centre, Cairo, Egypt\\
$^2$ Physical-Technical Institute, Tashkent, Uzbekistan\\
$^3$ P.N.Lebedev Physical Institute, Moskow, Russia\\
$^4$ Joint Institute for Nuclear Research, Dubna, Russia

\newpage

\section{Introduction}

Interesting experimental results on nuclear multifragmentation at low
and intermediate energies have recently been obtained. The radial
spherical symmetrical flow of fragments, or, in other words, the
proportionality of the kinetic energy of a fragment to its mass was
discovered in central collisions of gold projectile \cite{1} --
\cite{5}. The existence of the flow contradicts the assumption of a
stochastic character of the nuclear multifragmentation process. The
radial flow of spectator fragments in the rest frame of a fragmenting
nucleus has been found at study of projectile gold nuclei interactions
with photoemulsion nuclei at energy of 10.7 GeV/nucleon \cite{6, 7}.
According to estimations of Ref. \cite{5}, the energy of the radial
motion ranges between 30 \% and 50 \% of accessible energy. At the same
time, the another of the remarkable experimental result \cite{8}
obtained by the ALADIN collaboration -- the constancy of isotope
temperature over the wide range of the excitation energy, is treated as
an evidence of first order phase transition (liquid-gas) and
statistical nature of the nuclear multifragmentation. The third
experimental result \cite{9} reported by the INDRA collaboration -- the
independence of isotope composition of the fragments on the mass of
fragmenting nuclei at the same excitation energy, - is interpreted as a
sign of spinoidal instability of residual nuclei. There is a direct
evidence of a perplexing situation.

We believe that the present state of the activity may be settled by
studying the fragmentation of light enough systems (in the
investigations mentioned heavy systems were analyzed). In fact, the
famous statistical multifragmentation model (SMM) \cite{10, 11} was in
use to describe both heavy nuclei and oxygen residuals fragmentation
\cite{12}. No less famous the quantum molecular dynamic model
\cite{13}, in particular, the quantum antisymmetrized molecular
dynamics model \cite{14} which incorporates effects of the mean field,
are generally applied to light nucleus induced interactions. Hence,
there is an intersection region where two approaches can be applied.
The purpose of this work is to present the experimental data on
multifragmentation of residuals created in 0.9 GeV/nucleon krypton
initiated interactions in emulsion.

We start our analysis with the most interesting dependence of multiplicity
of intermediate mass fragments (IMF) on the mass of fragmenting system
whose measure is the so-called "bound" charge:
$$ Z_{bound} = \sum _F Z_F,~~~(Z_F \geq 2) $$
or
$$ Z_{b3} = \sum _F Z_F,~~~(Z_F \geq 3), $$
where $Z_F$ is a charge of a fragment. IMFs have charges
$3 \leq  Z_F \leq 30$.

Next, we consider intrinsic characteristics of the fragmenting systems
such as, the average charge of the largest fragment, asymmetry in the
fragment system, etc. Our consideration is summarized in a brief
summary.

\section{Experimental material}

Stacks of NIKFI BR-2 emulsion pellicles were exposed to a 1 A GeV
$^{84}Kr$ beam at the SIS/GSI  and a 10.7 A GeV  $ ^{197}Au$ beam at
the BNL/AGS. The sensitivity of the emulsion was not worse than 30
grains per 100 $\mu$k for singly charged particles with minimal
ionization.

To carry out the analysis, the events induced by krypton nuclei with
energy in the interval 0.8 -- 0.95 GeV/nucleon were taken. The mean
collision energy for this sample was thus reduced to about 0.9 GeV per
projectile nucleon.

All the interactions were found by along-the-track-"fast-slow" scanning
with a velocity excluding any discrimination in the event selection.
A slow scanning (in backward direction) was made to find the events
with unbiased projectile track. After excluding the events of
electromagnetic dissociation and purely elastic scattering, a total of
677 krypton-emulsion interactions and 1057 gold-emulsion interactions
heve been obtained.

Experimentally, the spectator fragments with $Z=2$ were classified by
the visual inspection of tracks. The ionization of such tracks is constant
over the whole length and equals $g/g_0= 4$, where $g_0$ is the
minimal ionization of singly charged track. Charge assignment for
multiply charged tracks were provided by delta-electron density
measurements on the length not less than 10 mm. The calibration
was made up on known primary tracks and tracks of the double charged
fragments. The accuracy of the measured charges was around three
charge units for $Z > 40$ and one charge unit for$ Z < 20$.

The relativistic particles emitted at $\theta < \theta_0$ ($\theta$
is the emission angle) were considered as singly charged fragments.
$\theta_0$ is determined as
$$ \sin \theta_0=0.2/P_0$$
where $P_0$ is the projectile nucleus momentum per nucleon in GeV/c.

In each investigated event, the polar $\theta$ and azimuthal $\varphi$
angles of all charged particles were measured.

The transverse momentum of a spectator fragment was defined as
\begin{equation}
\mid P_F \mid = 2 Z_F P_0 \sin \theta
\end{equation}
The ratio $A/Z$ for
fragments was assumed to equal two. The mean relative accuracy
in the transverse momentum of fragments does not exceed 7\%.

It should be noted that at high energies in contrast to low and
intermediate ones two clear-out distinguishable regions corresponding
to the projectile and target fragmentation are observed. The
probability of compound nucleus creation is assumed to be small. Thus,
the question how to select the fragments of the projectile and target
nuclei is simply settled.  Projectile fragments are regarded to have
the velocities equal to that of projectile nucleus. Having this in
mind, the relation (1) was suggested.  Clearly, it is invalid for deep
inelastic collisions where the fragments loss significant parts of
their longitudinal momentum. The lack of necessary experimental
information and specificity of the  photoemulsion experiments prevent
this circumstance from being taken into consideration.

The assumption of equality between the number of protons $P$ and
neutrons $N$ in the fragments brings main uncertainty to fragment
momenta.  For heavy nuclei $(N >P)$, the relation (1) underestimates
the transverse momentum. For double charged fragments among which
${}^3He$ isotopes are presented, the relation (1) overestimates, on
average, the transverse momentum. As shown in \cite{17}, 10\% ${}^3He$
admixture among all the fragments with $Z=2$ gives less than 1\%
growth in the dispersion of the transverse momentum (the dispersion
changes from 162 to 164 MeV/c). This result cannot, of course, have any
effect on the conclusions of the present paper.

Identification of the target fragments in photoemulsion experiments
requires a special track measurement technique that is not used at
our studies. Thus our data concern the fragments of
projectile nuclei, which were identified by the commonly accepted
emulsion analysis procedure. At photoemulsion studies dedicated to high
energies ($E >1$ A GeV), the projectile fragments are usually called
as spectator fragments. We will follow this tradition, sometimes
omitting the assignment "spectator".

\section{IMF multiplicity dependence on nuclear residual mass}

It is obvious that the multiplicity of intermediate mass fragments rises
with increasing the excitation energy at a fixed nuclear residual mass.
However, at high excitation energies, the production of light fragments
gets dominant, and $<N_{IMF}>$ must decrease. It is just the
dependence that has been observed by the ALADIN- group \cite{18} --
\cite{20} for multifragmentation of gold residuals created in the
interactions of the gold projectile with various targets at an energy
of 600 MeV/nucleon.  The "bound" charge $Z_{bound}$ which includes the
charges of $\alpha$ -particles  was taken as a measure of nuclear
residual mass.  Since $\alpha$-particles can be produced at the
pre-equilibrium decay stage, another value, $Z_{b3}$ \cite{21}, was
suggested to be used.  The most remarkable result was that $<N_{IMF}>$
as a function of $Z_{bound}$  or $Z_{b3}$ had proved independence of
the target mass.  As the dependence under discussion is essentially
determined by a relation between the excitation energy and the nuclear
residuals mass, one can conclude that the gold residuals of the same
masses from various reactions have nearly the same excitation energies.
Our results of Fig.1 allows one to make the statement more precise.

Fig.1 displays the $IMF$ multiplicity as functions of $Z_{bound}$ and
$Z_{b3}$ (closed circles) in the interactions of krypton nuclei with
photoemulsion nuclei at energy of 900 MeV/nucleon in a comparison with
the analogous data on 10.7 GeV/nucleon gold nuclei interactions (open
circles) \cite{6}. As seen, the data points for $Kr$ and $Au$ nuclear
residual fragmentation are close to each other for $Z_{bound}\leq 22$
and $Z_{b3}\leq 16$.  Although the error bars for $Kr$-points are quite
large, we can say that at the average $IMF$ multiplicity for $Kr$
projectile is larger than that for gold one at the same value of
$Z_{bound}$. At the same time, the  $<N_{IMF}>$ as functions of
$Z_{b3}$ for two projectiles are practically in a coincidence at
$Z_{b3} \leq 16$. This indicates that nuclear residuals of the same
masses formed in interactions of different systems at high energies
have approximately the same excitation energies if the initial nuclei
have lost more than one half of their nucleons at the fast stage of
the collisions.

Focus attention on fast growth of the our $IMF$ multiplicity in the
region of $Z_{b3}\sim 30$ with decreasing $Z_{b3}$. The rise seems to
be related to a threshold character of the nuclear multifragmentation.
Clearly, at small excitation energy the process of evaporation of
nucleons and light nuclei is dominated. At large excitation energy, the
multifragment decay channel opens. It was not clear whether the
probability of the last process evolves smoothly with excitation energy
increase or whether it is of threshold character. It is difficult to
note a change in the evolution of $IMF$ multiplicity at large $Z_{b3}$
for gold residuals fragmentation due to large error bars. The
statistics of the data is not rich enough to give conclusive results.
The data of the ALADIN collaboration at $Z_{b3} \geq 70$ have the
required statistics but, seemingly, suffer from methodical
uncertainties. We believe that a study of $IMF$ multiplicity at large
$Z_{b3}$ intended to look for the threshold character of the nuclear
multifragmentation is of great interest.

\section{Dependencies of intrinsic characteristics of decaying system
         on $Z_{bound}$ and $Z_{b3}$}

A quite unexpected result was obtained at an analysis of the heaviest
fragments. Fig.2 shows the mean charge of the heaviest fragment in
an event as a function of $Z_{bound}$ or $Z_{b3}$ (symbols are the
same as in Fig.1). One can see a clear change in the dependence of
$<Z_{max}>$ on $Z_{b3}$ at $Z_{b3} \sim  17$.  The analogous change,
but not so pronounced, has been found in the gold residuals
fragmentation.  This effect is slightly shaded when $Z_{bound}$ is
used. Thus we can conclude that nuclear residuals fragment in different
manners when more or less one half of nucleons from the primary nucleus
are ejected.  The data presented in Fig.3, where the average asymmetry
coefficient as a function of $Z_{bound}$ or $Z_{b3}$ is plotted,
confirms the above conclusion.

The value of $A_{12}$ for each event is determined as
$$A_{12}=\frac{Z_1 - Z_2}{Z_1 + Z_2},$$
where $Z_1$, $Z_2$, etc - the charges of fragments ordered such that
$Z_1 \geq Z_2 \geq Z_3$...,  $Z_1 \equiv Z_{max}$. As seen, there is
not any peculiarity in the dependence of $<A_{12}>$ on $Z_{bound}$.
$<A_{12}> $ as a function of $Z_{b3}$ remains practically constant at
$Z_{b3} \leq 17$ and then increases sharply with $Z_{b3}$ -- the
events with $Z_{b3}>17$ have strong asymmetry. A similar behavior is
also observed at the fragmentation of heavier systems.

Summing up, we can conclude that there exist at least two
regimes of fragmentation.

\section{Energy of fragments as functions of $Z_{bound}$ or $Z_{b3}$}

According to the statistical model of nuclear multifragmentation, the
kinetic energy of the fragments in the rest frame of fragmenting
nucleus is determined by the charge of the residual. Hence, the
decrease of the fragment energies can be expected with
decreasing $Z_{bound}$ \cite{22}. Going from the laboratory system
to a rest frame of a fragmenting nucleus, we have used the Gallilean
transformation described in Ref. \cite{6}. The average kinetic energy
of a fragment is connected with its transverse momentum, assuming the
isotropic decay, by
$$
<E> = (3/2)<P_F^2>/4 Z_F  m_N,
$$
Here, $P_F$ - the transverse momentum of the fragment, $Z_F$ - its
charge and $m_N$ - the mass of the nucleon.

Fig. 4 presents the mean kinetic energies of the fragments in the events
with the number of multiply charged fragments $(Z_F\geq 2)$ larger
or equal three. One can see that the kinetic energy of $Kr$ fragments
has no tendency to be decreased with decreasing the mass of residuals
in the region $Z_{bound} \leq  25$. Moreover, in the region they are
practically permanent. A similar behavior is observed at the
fragmentation of gold residuals. The results provide an evidence for
the radial flow of the fragments, with the energy of the flow depending
on the mass of the initial nucleus (according to our data).

\section{Summary}
The experimental data on the multifragmentation of the krypton
residual nuclei formed in the interactions with photoemulsion nuclei at
energy 0.9 GeV/nucleon are presented.

The mechanism of the nuclear residual fragmentation is shown to be
practically independent on the mass of projectile nucleus if
$Z_{bound}$ does not exceed a one half of the charge of the initial
nucleus.

The evidences are obtained that the multifragmentation is of
a threshold character and that there is a radial flow of the
fragments that depends upon the mass of the initial nucleus.

The experimental regularities manifest themselves more brightly when
the value $Z_{b3}$ is used as a measure of the residual mass.

Further experimental and theoretical studies of the
multifragmentation of intermediate mass nucleus are of interest.

The authors are thankful to the EMU-01 collaboration for providing
us with the experimental material of the gold interactions.
The financial support from
the Russian Foundation of Fundamental Research (grant No. 99-02-17757),
and
the Foundation of Fundamental Research of Academy of Sciences of Uzbekistan,
are cordially and gratefully acknowledged.

\newpage 

\begin{center}
 Figure captions
\end{center}

\noindent Fig.1: The dependencies of the $IMF$ multiplicity on
$Z_{bound}$ (a) and $Z_{b3}$ (b). Closed circles - data for $Kr$
interactions at E=900 MeV/nucleon, open circles - data for $Au$
interactions at E= 10.7 GeV/nucleon.

\vspace{1.5cm}
\noindent Fig.2: The average charge of the largest  fragment in the
events as a function of "bound" charge. Symbols are the same as in
Fig.1.

\vspace{1.5cm}
\noindent Fig.3: The average coefficient of asymmetry in the measured
events as a function of "bound" charge. Symbols are the same as in
Fig.1.

\vspace{1.5cm}
\noindent Fig.4: The estimated average kinetic energy of the
fragments in the rest frame of the fragmenting nucleus. Symbols are the
same as in Fig.1. a) - energies of double charged fragments; b)
 - energies of the fragments with $Z_F=3-5$; c) - energies of $IMF$s.

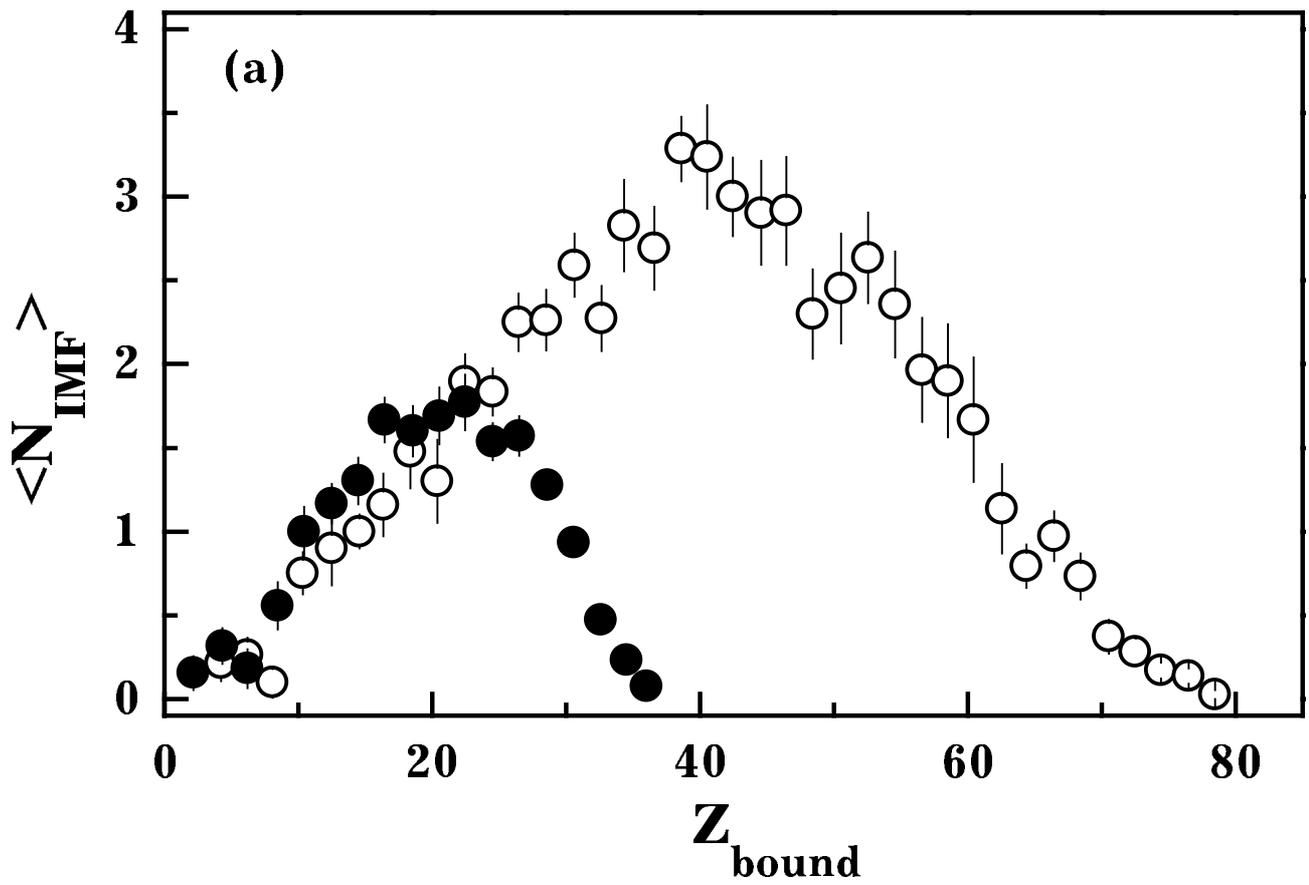
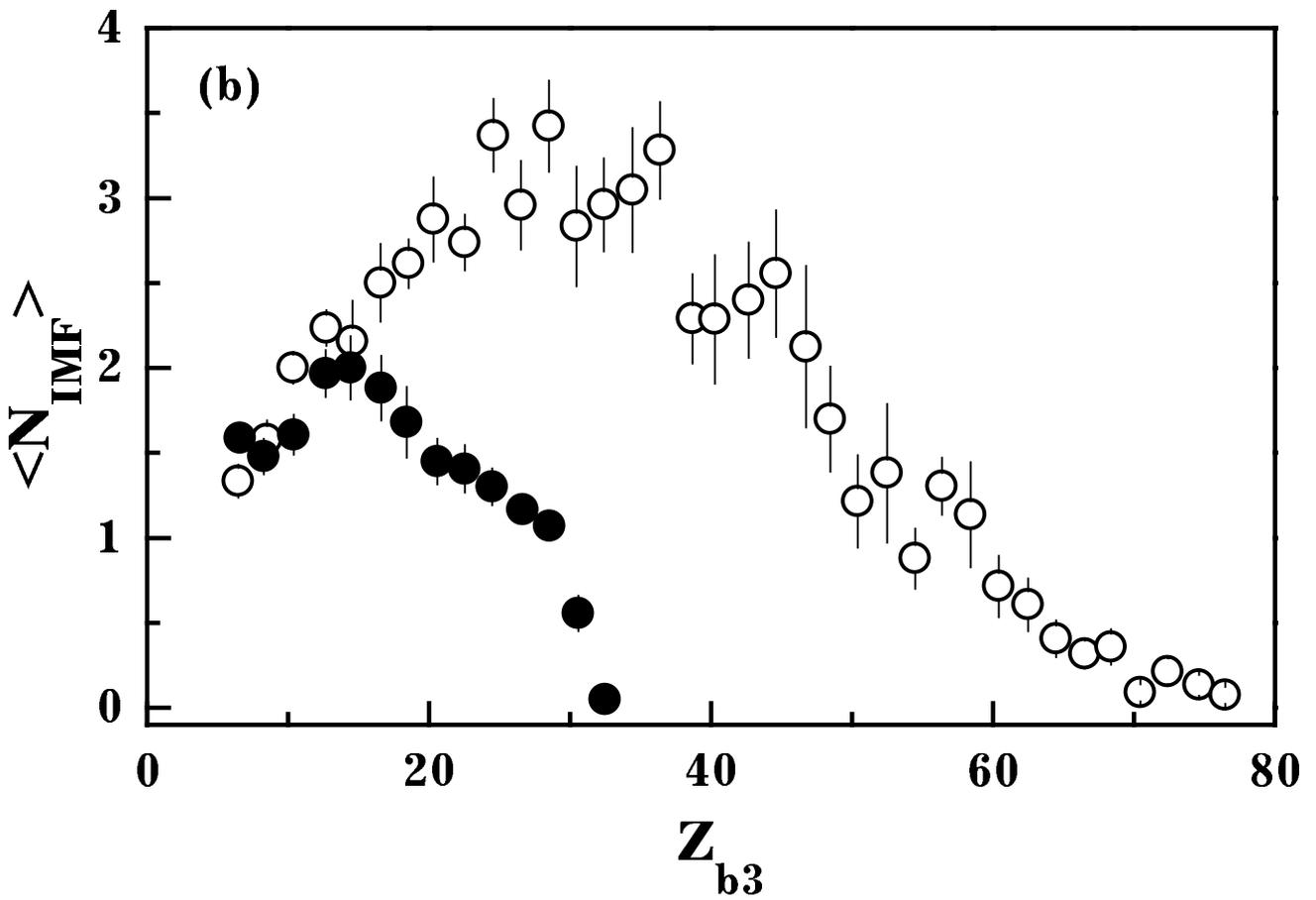
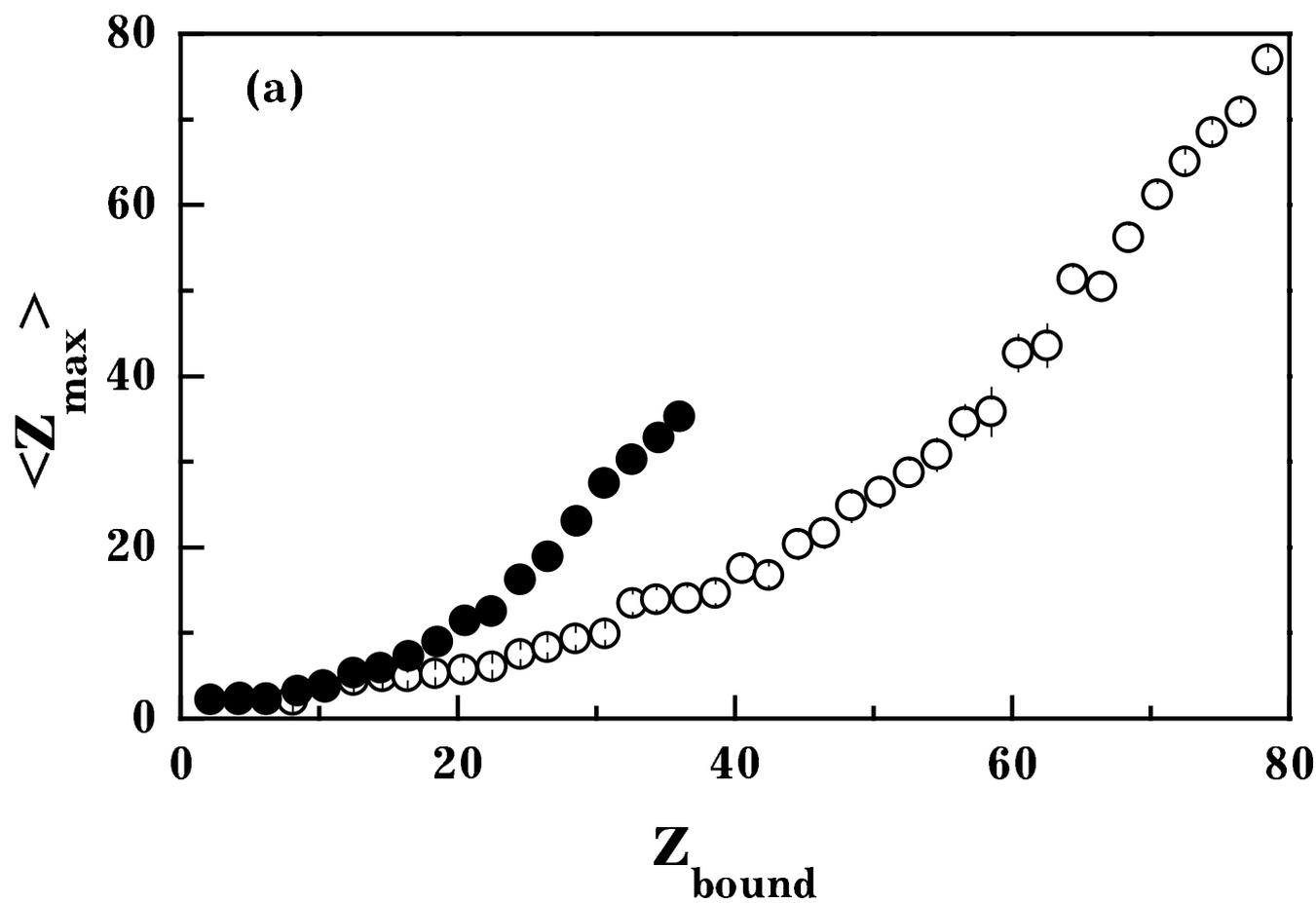
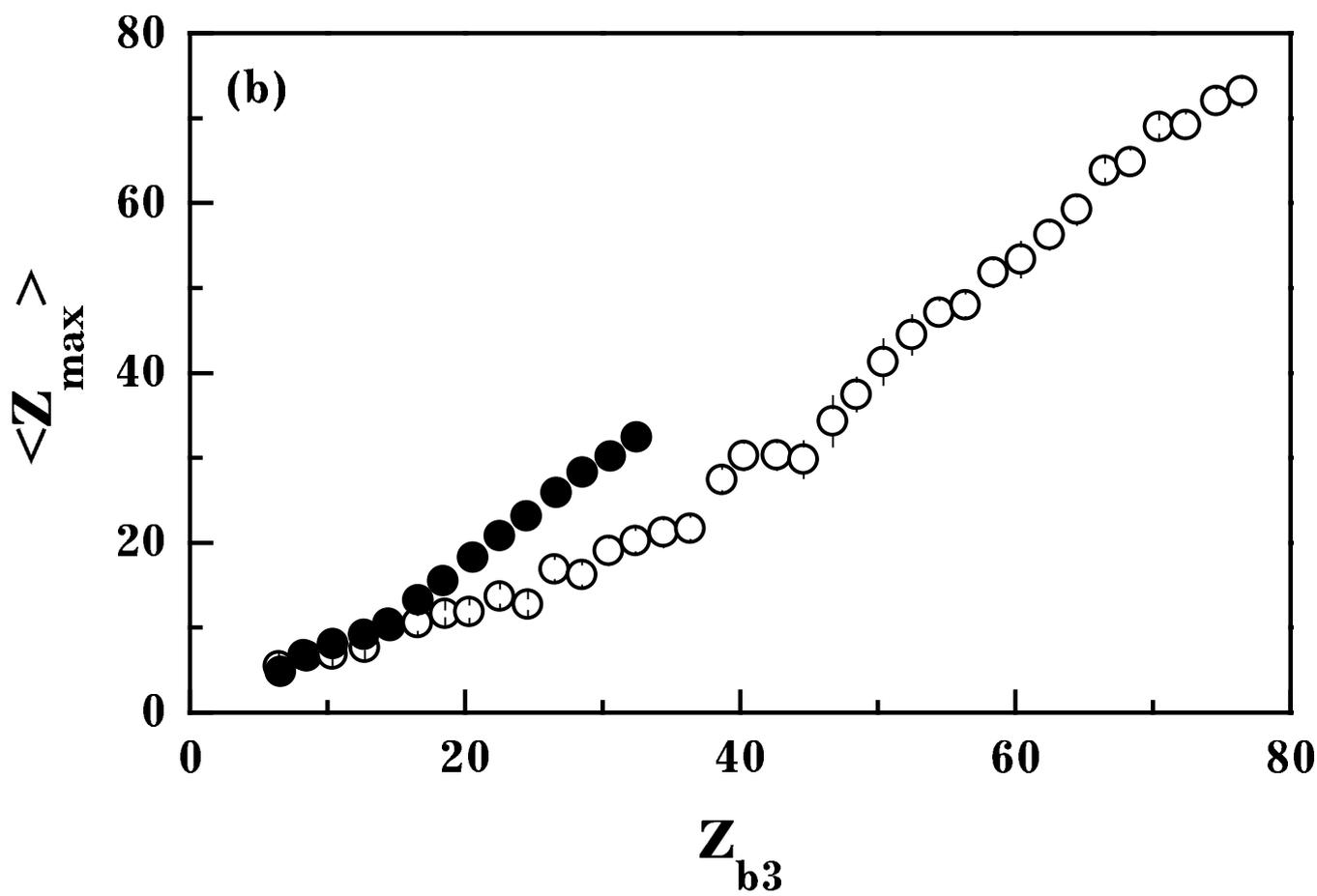
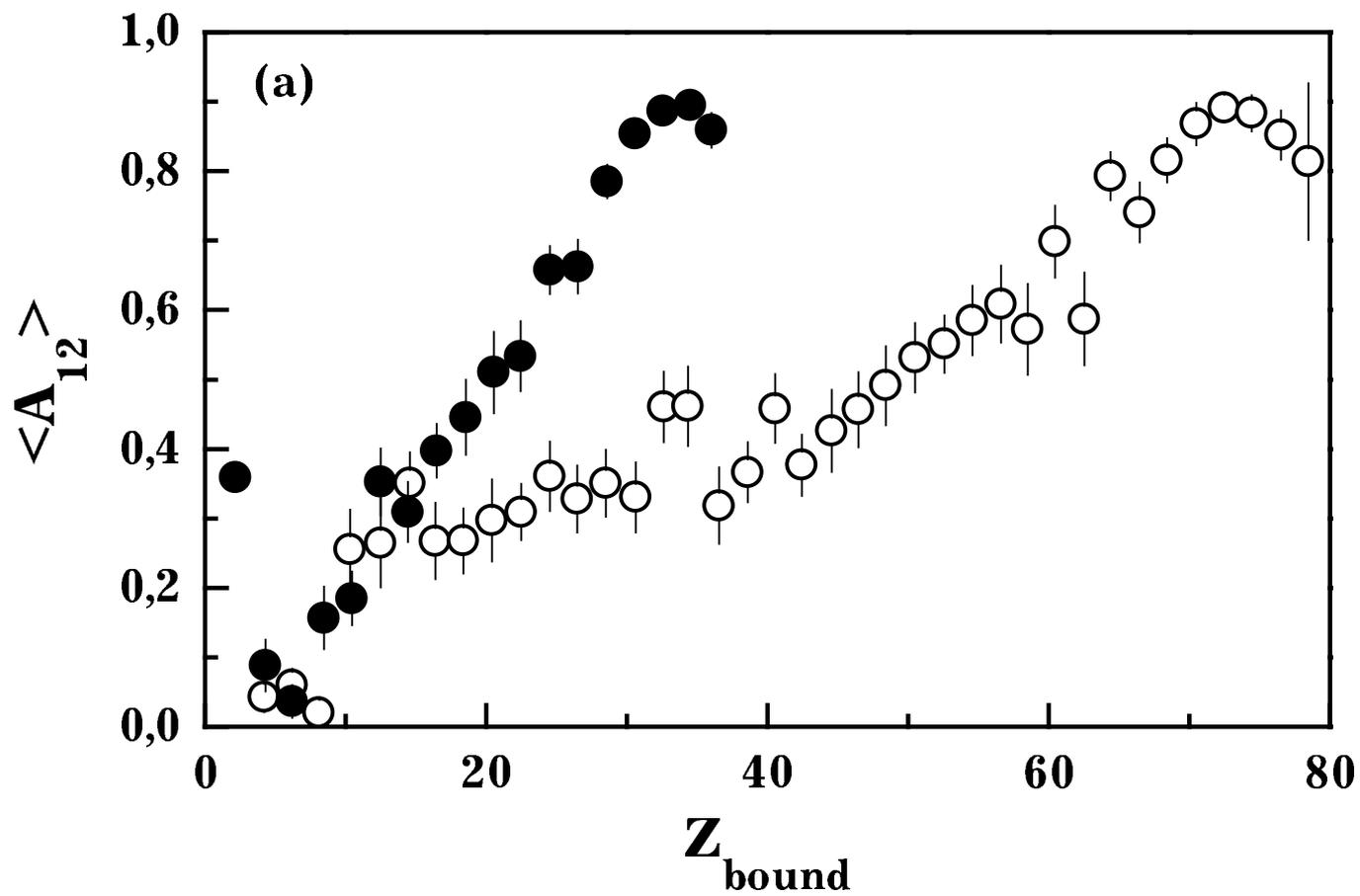
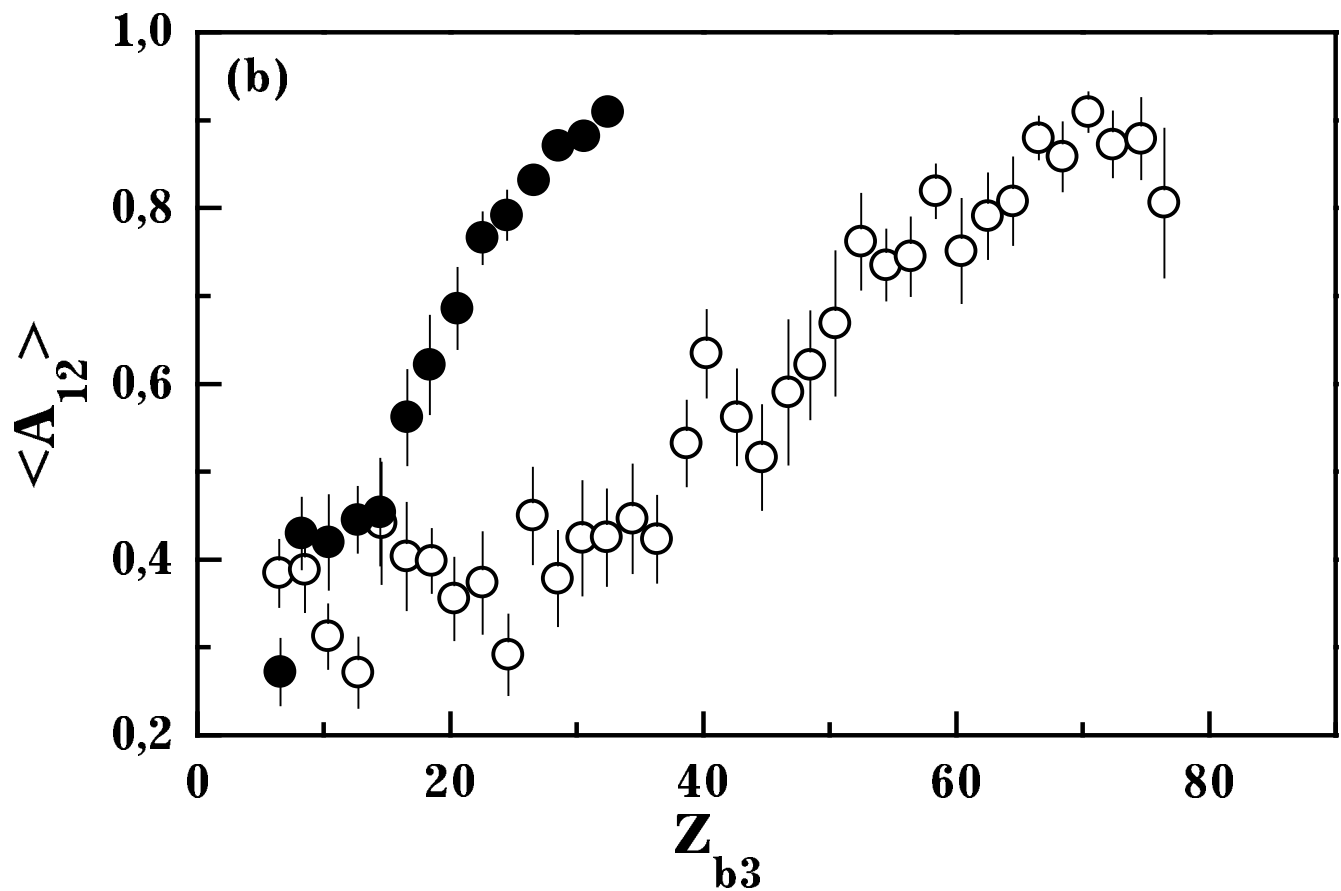
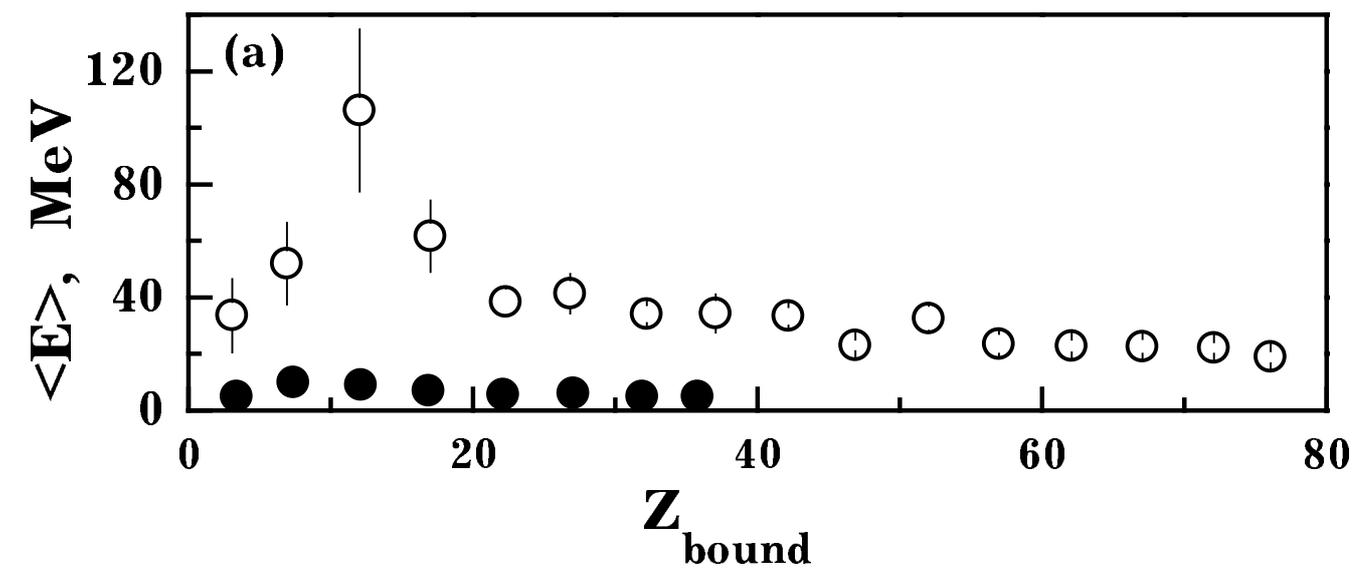
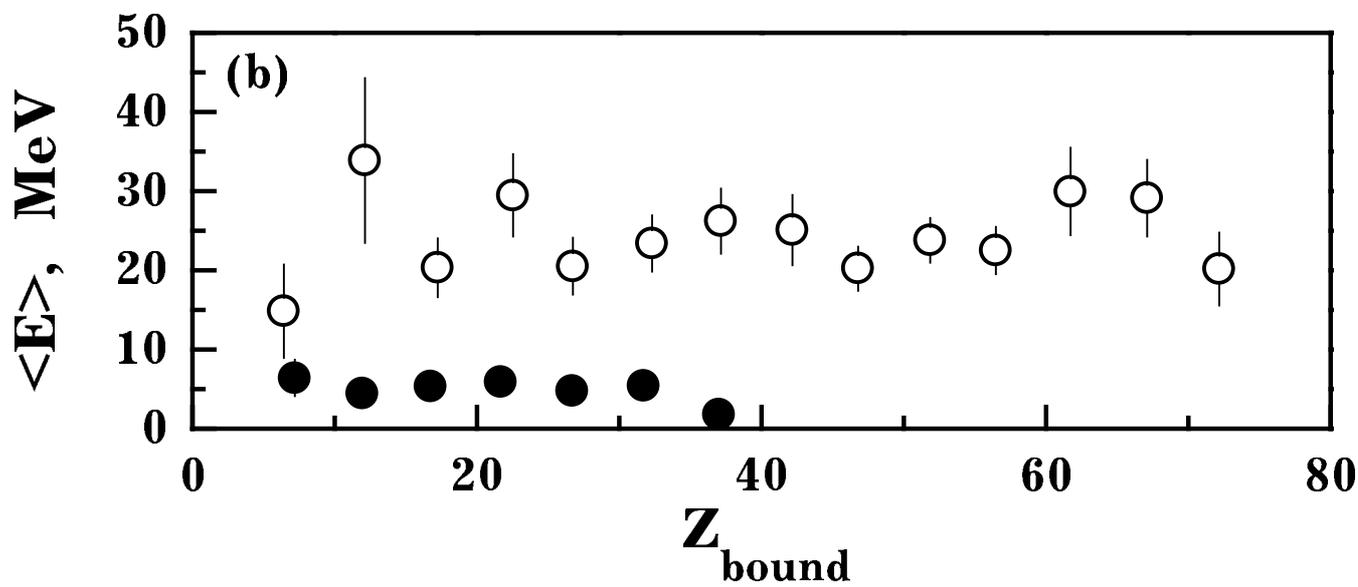
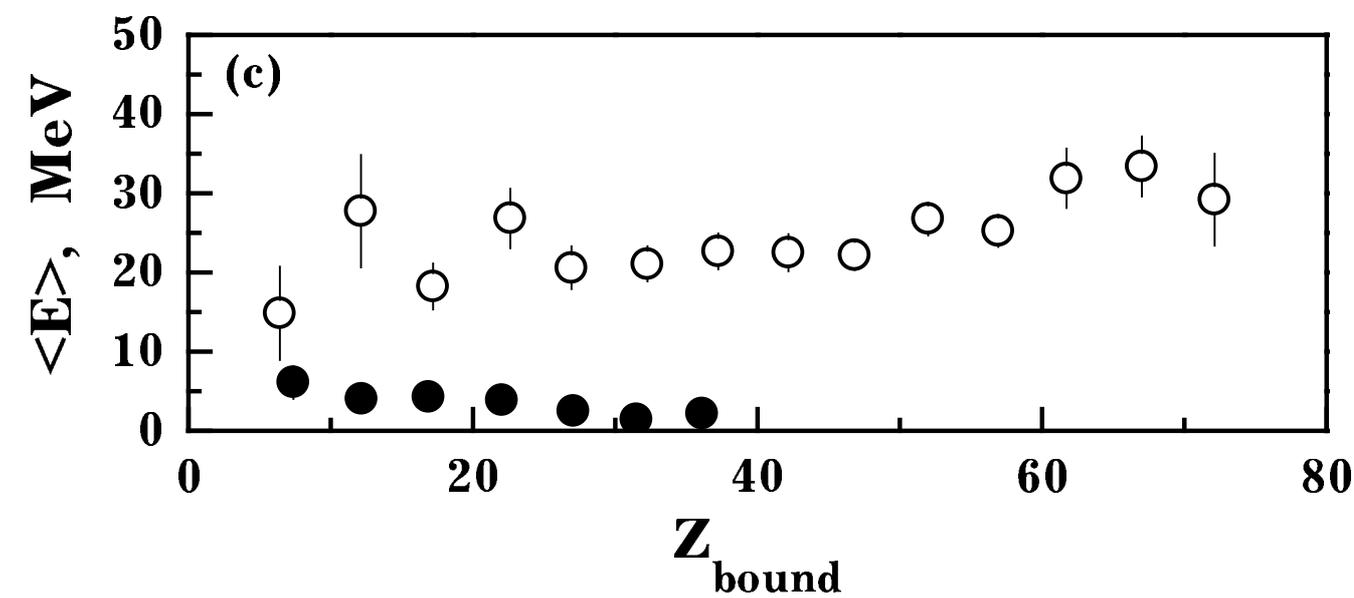
\end{document}